\documentclass[12pt]{article}
\usepackage{amsfonts}
\usepackage{color}
\usepackage[active]{srcltx}

\begin{document}

\def\ba{\begin{eqnarray}}
\def\ea{\end{eqnarray}}
\def\w{\wedge}

\begin{titlepage}
\title{ \bf Non-minimal $RF^2$-type corrections to holographic superconductor}
\author{ \"{O}zcan Sert$^{a}$\footnote{\tt osert@pau.edu.tr} \hskip 0.5cm and \hskip 0.3cm Muzaffer Adak$^{b}$\footnote{\tt madak@pau.edu.tr}\\ \\
 {\small $^a$Department of Mathematics, Faculty of Arts and Sciences, Pamukkale University}\\
 {\small 20017 Denizli, Turkey }\\
 {\small $^b$Department of Physics, Faculty of Arts and Sciences, Pamukkale University}\\
 {\small 20017 Denizli, Turkey }   }
 \vskip 1cm
\date{22 October 2013 {\it file:RF2holsupercond04.tex}}

\maketitle

\begin{abstract}

\noindent We study $(2+1)$-dimensional holographic superconductors
in the presence of  non-minimally coupled electromagnetic field to
gravity by considering an arbitrary linear combination of
$RF^2$-type invariants with three parameters. Our analytical
analysis shows that the non-minimal couplings affect the
condensate and the critical temperature.

\vskip 1cm


% 11.25.Tq    Gauge/string duality
% 03.50.De    Classical electromagnetism, Maxwell equations (for applied classical electromagnetism, see 41.20.-q)
% 04.50.Kd    Modified theories of gravity

\end{abstract}
\end{titlepage}

\def\ba{\begin{eqnarray}}
\def\ea{\end{eqnarray}}
\def\w{\wedge}

\section{Introduction}

Anti-de Sitter/conformal field theory (AdS/CFT) duality emerged
from string theory has given rise to novel deductions on the
research of superconductivity in condensed matter physics.
According to AdS/CFT dictionary a suitably chosen gravitational
theory in four dimensions (bulk) can describe basic properties of
a superconductor in three dimensions (boundary). Holographic
superconductors have been worked extensively in literature, see
some selected papers \cite{gubser}-\cite{Roychowdhury} and
references therein. Although in this regard generally numerical
solutions were investigated, endeavors of finding analytic
solutions occurred after the paper \cite{Gregory} in which the
authors found that higher curvature corrections make condensation
harder by using an analytic approximation method.

In Ref.\cite{Sonner}, based on a numerical approach, the author
found that the critical temperature depends on the rotation by
applying the method developed by Hartnoll et al \cite{Hartnoll} to
a rotating holographic superconductor. In Ref.\cite{Pan} the
authors considered the Maxwell field strength corrections by
following the technique in \cite{Sonner} and concluded that the
higher correction to the Maxwell field makes the condensation
harder to form. In Ref.\cite{Roychowdhury}, based on an analytic
method, the author investigated several properties of holographic
superconductors by incorporating separately the Born-Infeld term
and the Weyl-invariant correction term to the standard bulk
lagrangian. As distinct from those, in this Letter we aim to
analyze the effects of non-minimal $RF^2$-type contributions to
the standard holographic superconductor lagrangian. Similar terms
have appeared in a calculation in QED of the photon effective
action from 1-loop vacuum polarization on a curved background
\cite{drummond}, and in the Kaluza-Klein reduction of $R^2$
lagrangian from five dimensions to four dimensions \cite{dereli1}.
Furthermore, as the relevances of such terms to the dark matter,
the dark energy, the primordial magnetic fields in the universe
have been investigated in the references
\cite{lambiase}-\cite{dereli4}, the gravitational wave concerns
have been discussed by searching pp-wave solutions in
\cite{dereli2}. Therefore, it is worthwhile to work possible
effects of $RF^2$-terms on holographic superconductor. In fact,
some specific combinations of that type interactions have appeared
in previous holographic studies. For example, in
Ref.\cite{myers2011} the authors have investigated the effects on
the charge transport properties of the holographic CFT resulting
from the extra four-derivative interaction formulated in terms of
the Weyl tensor which is constructed as a particular linear
combination of our $c_1,c_2,c_3$ terms in the lagrangian density
(\ref{lagrange}). Thus, one novelty of this work is to keep three
(perturbatively small) unspecified coupling constants $c_{1,2,3}$
independent.

We work  in the probe limit in which the electromagnetic field and
scalar field do not backreact  on the geometry, and use the
analytic approximation method developed by Gregory et al in
\cite{Gregory}. Their method explains the qualitative features of
superconductors and expects quantitatively accurate numerical
results. Consequently, we obtain an analytic expression for the
condensate and the critical temperature in the existence of the
non-minimal couplings. Correspondingly, we observe that they are
are affected critically by the non-minimal coupling parameters.

\section{The Model} \label{model}

The Riemannian bulk spacetime is denoted by $\{M,g\}$ where $M$ is
four-dimensional differentiable and orientable  manifold endowed
with a non-degenerate metric $g$. We will be using orthonormal
1-form $e^a$ such that $g=\eta_{ab} e^a \otimes e^b$ where
$\eta_{ab}=\mbox{diag}(-1,1,1,1)$. Orientation is fixed through
the Hodge map $*$ such that $*1 = e^0 \w e^1 \w  e^2 \w e^3$ where
$\wedge$ figures the exterior product. The Riemann curvature
2-form is defined by $R^a{}_b = d\omega^a{}_b + \omega^a{}_c \w
\omega^c{}_b$ where $\omega_{ab}=-\omega_{ba}$ is the Levi-Civita
connection 1-form $\omega^a{}_b \w e^b = -de^a$. We will use the
following shorthand notations throughout the Letter; $ e^a \wedge
e^b \wedge \cdots = e^{ab\cdots}$, $\iota_a \iota_b \cdots =
\iota_{ab\cdots}$, $\iota_aF =F_a$, $\iota_{ba} F =F_{ab}$, $
\iota_a {R^a}_b =R_b$, $ \iota_{ba} R^{ab}= R $ where $\iota_a$
denotes the interior product such that $\iota_b e^b=\delta^a_b$.
Here $\delta^a_b$ is the Kronecker symbol.

We consider the following lagrangian density 4-form
 \begin{eqnarray} \label{lagrange}
  \mathcal{L} &=&\frac{1}{\kappa^2}\left[ R_{ab} \w *e^{ab}+ \Lambda *1
                   + (de^a + \omega^a{}_b \w e^b)\w \lambda_a \right] \nonumber \\
        & &-\frac{1}{2} F\w *F   -D\psi^{\dag}\wedge D\psi -m^2\psi^{\dag}\psi*1  \nonumber \\
    & &  + \frac{c_1}{2} F^{ab} R_{ab} \w *F + \frac{c_2}{2} F^a\wedge R_a \w *F   + \frac{c_3}{2} RF\w
    *F
   \end{eqnarray}
where $\kappa$ is the gravitational coupling coefficient, $F=dA$
is the Maxwell 2-form with the electromagnetic potential 1-form
$A$, $\Lambda$ is the cosmological constant, $\psi$ is the complex
scalar field (hair), the dagger symbol signifies complex
conjugation, $m$ is the mass of hair, $c_i$, $i=1,2,3,$ are
non-minimal coupling coefficients, $\lambda_a$ is lagrange
multiplier constraining connection to be Levi-Civita and $ D\psi =
d\psi + i A \psi$. The first two lines of (\ref{lagrange}), i.e.
the choice of $c_1=c_2=c_3=0$, is the standard holographic
superconductor lagrangian introduced by Gubser in \cite{gubser}.
The third line is a linear combination of three non-minimal
$RF^2$-type terms. The case for vanishing hair and cosmological
constant has been worked much for various reasons such as the dark
matter, the dark energy and the primordial magnetic fields in the
universe, the gravitational waves \cite{drummond}-\cite{dereli4}.
Another special choice in which the $RF^2$-lagrangian is
conformally invariant is $c_3=\gamma/3$, $c_1=c_2=\gamma$. Here it
is also important to note that these non-minimal terms are taken
to be perturbative, otherwise the model suffers from a number of
problems, e.g. the presence of ghosts. Therefore we consider  the
perturbatively small coefficients $c_i$.

Now, since we will be working in the probe limit, as usual we
concentrate only on the electromagnetic field equation and the
hair field equation obtained by independent variations with
respect to $A$ and $\psi^\dag$, respectively,
\begin{eqnarray}
 d\Big\{-*F  +\frac{c_2}{2}\big[ R_a\wedge \imath^a*F-R*F+*(F^a\wedge R_a)\big]& & \label{Maxwellfe} \\
  + c_1*F^{ab}R_{ab} + c_3R*F \Big\}- 2|\psi|^2*A &=& 0,\nonumber \\
   D*D\psi -m^2\psi*1 &=& 0 \, . \label{scalarfe}
  \end{eqnarray}
We can assume that $\psi$ is everywhere real and the mass is
$m^2=-2/L^2$ which satisfies the Breitenlohner-Freedman bound,
$m^2L^2 \geq -9/4$. In order to find a solution to those equations
we start with the metric of a planar Schwarzschild-AdS black hole
 \begin{equation}
    g = -f(r)dt^2  +  \frac{dr^2}{f(r)} + \frac{r^2}{L^2}(dx^2 + dy^2)
 \end{equation}
where
  \begin{equation}
            f(r) =\frac{r^2}{L^2}(1-\frac{r_H^3}{r^3}) \, .
   \end{equation}
Here we write the cosmological constant in terms of the AdS radius
$L$ as $\Lambda=6/L^2$ and the mass of the black hole $M$ in terms
of the position of the horizon $r_H$ as $M= r_H^3/L^2$. Now we
think of the case $\psi=\psi(r)$ and $A=\phi(r)dt$. Then the
passage to a new independent variable through $z=r_H/r$ brings the
outer region $r_H \leq r \leq \infty$ to the interval $0 \leq z
\leq 1$. Correspondingly, the equations (\ref{Maxwellfe}) and
(\ref{scalarfe}) turn out to be
 \begin{eqnarray}
    \left[1 + \frac{2c_1}{\beta L^2}(1-z^3)\right]\phi_{zz}  - \frac{6c_1}{\beta L^2}z^2\phi_z
    - \frac{2L^2}{\beta}\frac{ \psi^2}{ z^2(1-z^3)} \phi =  0 \, ,\label{phi1}\\
     \psi_{zz}- \frac{2+z^3}{z(1-z^3)}\psi_z + \left[\frac{L^4\phi^2}{r_H^2(1-z^3)^2}  +
         \frac{2}{z^2(1-z^3)}\right]\psi = 0 \, , \label{psi1}
 \end{eqnarray}
where $\beta= 1-6c_2/L^2 +12c_3/L^2$ and a subindex $z$ denotes
$d/dz$. Thus, we continue tracking $\beta$ apart from $c_1$ in
order to see the novel $RF^2$-contributions.

Now we enumerate the steps that we will pursue. Firstly, we make a
usual solution ansatz in the AdS region, $z\rightarrow 0$,
   \begin{eqnarray}
      \phi(z)=\mu - q z \, , \quad  \psi(z)=\psi_1z+\psi_2z^2 \, .
      \label{solasy}
   \end{eqnarray}
$\psi_1$ and $\psi_2$ are interpreted as condensates, and $\mu$
and $q$ as the chemical potential and the charge density,
respectively, in the dual theory. According to \cite{gubser}, for
$-9/4 < m^2L^2 < -5/4$ one can choose either $\psi_1=0$ or
$\psi_2=0$. So we take safely $\psi_1=0$. Secondly, we find an
approximate solution at the horizon, $z=1$, by using the Taylor
expansion technique and then apply the regularity condition at the
boundary
 \begin{eqnarray}
     \phi(1)=0 \, , \quad  \psi_z(1)=\frac{2}{3}\psi(1) \, ,
     \label{boundary}
   \end{eqnarray}
Thirdly, we match two solutions at an intermediate point $0 < z_m
< 1$. Finally we deduce $\psi_2$ and the critical temperature, and
comment on them.

Since the first step is already there, let us start with the
second step by writing down the Taylor expansion of $\psi(z)$ and
$\phi(z)$ around $z=1$
  \begin{eqnarray}
     \phi(z)=\phi(1)-\phi_z(1)(1-z)+ \frac{1}{2}\phi_{zz}(1)(1-z)^2+ \cdots \, ,\\
    \psi(z) =\psi(1)-\psi_z(1)(1-z)+\frac{1}{2}\psi_{zz}(1)(1-z)^2+ \cdots \, .
\end{eqnarray}
We calculate $\phi_{zz}(1)$ from (\ref{phi1}) and $\psi_{zz}(1)$
from (\ref{psi1})
  \begin{eqnarray}
  & &  \phi_{zz}(1)= \frac{6c_1}{\beta L^2}\phi_z(1)-\frac{2}{3\beta}L^2\phi_z(1)\psi(1)^2 \, ,\\
  & &  \psi_{zz}(1)=-\frac{2}{3}\psi_z(1) - \frac{L^4}{18r_H^2}\phi_z(1)^2\psi(1) \, .
\end{eqnarray}
By substituting these into above and also incorporating
(\ref{boundary}) we obtain a set of two serial solutions at the
horizon
  \begin{eqnarray}
 & &  \phi(z)=-\phi_z(1)(1-z) + \left[ \frac{3c_1}{\beta L^2}
             - \frac{1}{3\beta}L^2\psi(1)^2 \right] \phi_z(1)(1-z^2)+ \cdots \, , \label{phi2}\\
 & &   \psi(z)=\frac{1}{3}\psi(1)+ \frac{2}{3}\psi(1)z-\frac{2}{9}
             \left[ 1+\frac{L^4}{8r_H^2}\phi_z(1)^2 \right] \psi(1)(1-z^2)+   \cdots \,
             . \label{psi2}
\end{eqnarray}

Thus we are in the third step in which we equate $\phi(z)$ and
$\psi(z)$ in (\ref{solasy}) to (\ref{phi2}) and to (\ref{psi2}),
respectively, at the intermediate point $z_m$. Since allowing
$z_m$ to be arbitrary does not alter quantitative behaviors of the
analytic method \cite{Gregory}, we fix it as $z_m=1/2$. Smooth
matching yields four conditions
 \begin{eqnarray}
   \mu -\frac{q}{2} &=& \left(\frac{1}{2}-\frac{3c_1}{4\beta L^2} \right)b + \frac{L^2}{12\beta}ba^2 \, ,\\
    -q &=&\left(\frac{3c_1}{\beta L^2}   - 1 \right)b  - \frac{L^2}{3\beta}ba^2 \, ,\\
    \frac{\psi_2}{4} &=& \frac{11}{8}a - \frac{L^4}{144r_H^2}ab^2 \, ,\\
    \psi_2 &=& \frac{8}{9}a + \frac{L^4}{36r_H^2}ab^2 \, ,
\end{eqnarray}
where we have renamed $\psi(1)\equiv a $ and $-\phi_z(1)\equiv b$
($a,b>0$) for plain. These equations yield
  \begin{eqnarray}
    a^2 = \frac{3q\beta }{bL^2} \left(1- \alpha^2 \frac{b}{q}\right) \, , \quad
    \psi_2 = \frac{5}{3}a \, , \quad
     b = \frac{2\sqrt{7}r_H}{L^2} \, ,
 \end{eqnarray}
where $\alpha^2 = 1- 3c_1/\beta L^2$. From now on we need to track
$\alpha$ and $\beta$ for our novel results coming from
$RF^2$-terms. By fixing the charge density $\rho=qr_H$ and using
the Hawking temperature $T=3r_H/(4\pi L^2)$, we calculate the
expectation value of the dimension 2 operator $\langle
\mathcal{O}_2 \rangle= \sqrt{2}\psi_2r_H^2/L^3$ as
 \begin{eqnarray}
    \langle \mathcal{O}_2 \rangle = \frac{80\pi^2}{9}\sqrt{\frac{2\beta}{3}}TT_c
        \sqrt{1+\frac{T}{T_c}}\sqrt{1-\frac{T}{T_c}}\\
  \end{eqnarray}
where we defined the critical temperature
  \begin{eqnarray}\label{Tc0}
    T_c=\frac{3\sqrt{\rho}}{4\pi L \alpha\sqrt{2\sqrt{7}}} \, .
  \end{eqnarray}
For the case  $ \alpha \rightarrow 0 $  or $c_1 / \beta
\rightarrow L^2/3$ the critical temperature $T_c$ goes to
infinity. This case corresponds to having the higher order
corrections of the same order as the leading order result, which
can not be admissible for arising  ghosts  in such case. Hence, we
perturbatively expand  the term $1/\beta $   in $\alpha^2$  in
terms of the  small coefficients $c_i$ up to leading order.
 \begin{eqnarray}
\alpha^2 = 1-3c_1/L^2 + \mathcal{O}(c_i^2)
\end{eqnarray}
Then, the critical temperature $T_c$ is affected from the non-minimal correction in the leading order as follows,
    \begin{eqnarray} \label{Tc}
    T_c =\frac{3\sqrt{\rho}}{4\pi L \sqrt{2\sqrt{7}}} ( 1+ \frac{3c_1}{2L^2 }  )  +  \mathcal{O}(c^2_i) \, .
  \end{eqnarray}
Thus, $T_c$  increases as  $c_1 $ increases from zero to  $c_1 =
L^2/3$.

\section{Concluding remarks}

As expected, the condensation occurs below the critical point
$T_c$ and the mean field theory result $ \langle \mathcal{O}_2
\rangle \propto (1-T/T_c)^{1/2}$ is valid. Since the condensate is
proportional to $\sqrt{\beta}$, the consistency condition $\beta >
0$ causes a restriction between the non-minimal coupling constants
$c_2$ and $c_3$. The condensation is low for $\beta < 1$ and the
reverse is hold for $\beta > 1$. Besides, if the special case $c_2
=2c_3$ is encountered, $\beta$ goes to unity which means that the
condensate is not influenced.

We notice that for the special case $ c_1 = \beta L^2/3 $  in
(\ref{Tc0}) the critical temperature $T_c $ goes to infinity. This
case corresponds to having the higher order corrections of the
same order as the leading order result, which should not be
admissible for arising ghosts  in such cases. In order to avoid
such problems we expand the temperature in terms of $c_i$. From
(\ref{Tc}) we see that the critical  temperature depends  on $c_1$
in the leading order. We see that as $c_1$ increases,  $T_c$
increases. This case allows very high temperature superconductor.
But, there is  a constrain on the coupling parameter which is $c_1
\leq  L^2/3 $ in the leading order.

We notice also that for the conformally invariant case
$c_1=c_2=\gamma$ and $c_3=\gamma /3$ the Weyl parameter must
respect an upper bound $\gamma \leq L^2/5$ for a non-zero
condensate. This result is in favor of \cite{Roychowdhury} and
\cite{myers2011} in which certain aspects of the Weyl corrections
to holographic superconductor in a five and a four dimensional
bulk space-times have been discussed, respectively.

\section*{Acknowledgement}

\noindent One of the authors (\"{O}.S.) is  supported by the
scientific research project (BAP) 2012BSP014, Pamukkale
University, Denizli, Turkey.

\end{document}